# A New Mattress Development Based on Pressure Sensors for Body-contouring Uniform Support

Hsiu-Chen Hsu
Institute of Computer and Communication Engineering
National Taipei University of Technology
Taipei , Taiwan
c41sheu@yahoo.com.tw

Rong-Chin Lo
Institute of Computer and Communication Engineering
National Taipei University of Technology
Taipei , Taiwan
rclo@ntut.edu.tw

***Abstract*—For getting good sleep quality, an improved approach of new mattress development based on the pressure sensors for body-contouring uniform support is proposed in this paper. This method solved the problems of innerspring mattresses that cannot allow body-contouring uniform support, and foam mattresses that cannot provide everybody equal comfort from the same mattress. By the buried pressure sensor array and actuator array in foam layer of a mattress, both are connected to a controller to generate the pressure distribution mapping of a human body on the mattress, then from the data of this mapping, some of the actuators are driven up or down by the controller to generate a body-contouring uniform support. By the aid of mathematical morphology algorithms, user can also choose a different support mode by another wireless controller with touch-screen to accommodate personal favorite firmness of the mattress and to take his tensed mood and pressure off with good sleep until daylight. Moreover, some other homecare functions, such as temperature measurement, sleep on posture correction and fall down prevention, can approach by additional hardware and software as user requirement in the future.**

## I. INTRODUCTION

Chronic insomnia will cause physiological or psychological illness such as back pain, tendonitis, scoliosis and depression [1]. Except for a variety of pressures, wrong sleep style and environment are important ingredients for insomnia. To solve the latter, the function of mattress is most important. There are many kinds of mattress in the market. However, modern mattresses usually contain either an innerspring core or materials such as latex, viscoelastic or other flexible polyurethane foams. Innerspring mattresses with up-down resilient cannot allow body-contouring uniform support and will cause pressure concentrated to some points of the body such as shoulders, hips and heels. This uneven pressure, muscle will be tensed and increase turning frequencies when sleep. Foam mattresses with good resilience foams, softness and ventilation, conform to ergonomics. These foam mattresses can fit body curve and provide properly support and comfort. However, these mattresses cannot allow everybody equal comfort from the same mattress.

In this research, we proposed a method to keep the advantage of foam mattresses that can conform to ergonomics and to improve the innerspring mattress's disadvantages that lead body-contouring to non-uniform support. The best mattress should uniformly support the human body at all points, i.e. the body should be supported from traditional resilience support to full body uniform support. To do this, pressure sensor array are used to sense pressure of full body on the mattress and actuator array are used to provide support for each point of body. By buried these two components in foam layer of mattress, both are connected to a FriendlyARM microcontroller, and the pressure distribution mapping of full body on a mattress can be generated by the controller automatically. Then from the data of pressure distribution mapping, the weight of human body and uniform support can be calculated. Some actuators are driven up or down by the controller to provide a body-contouring uniform support, i.e. making even pressure for each support point of the human body. Additionally, by the aid of mathematical morphology algorithms [2]-[3]—to analyze the area of body morphology, user can choose a firmness level via a wireless controller with touch-screen to accommodate personal favorite firmness of the mattress and then to take his tensed mood and pressure off with good sleep until daylight.

## II. OVERVIEW OF THE PROPOSED SENSOR-BASED MATTRESS ARCHITECTURE

As can be seen in Fig. 1, the proposed architecture is mainly composed of four blocks, the sensor array on the top left, the actuator array on the bottom left, the bedding controller in the middle and the wireless controller with touch-screen on the right side. Sensor array are used to sensing the pressure of the human body on the mattress at all points. The sensed voltage signals are then amplified and processed through the A/D port of MCU. Actuator array are used to vary each point support of the mattress by up (CW) or down (CCW) driven through stepping clock. All support functions of the mattress are controlled by software embedded in MCU and hardware drivers. To bridge the user and the proposed sensor-based mattress, a wireless controller is required. User can monitor mattress status and transmit commands via ZigBee protocol [4].

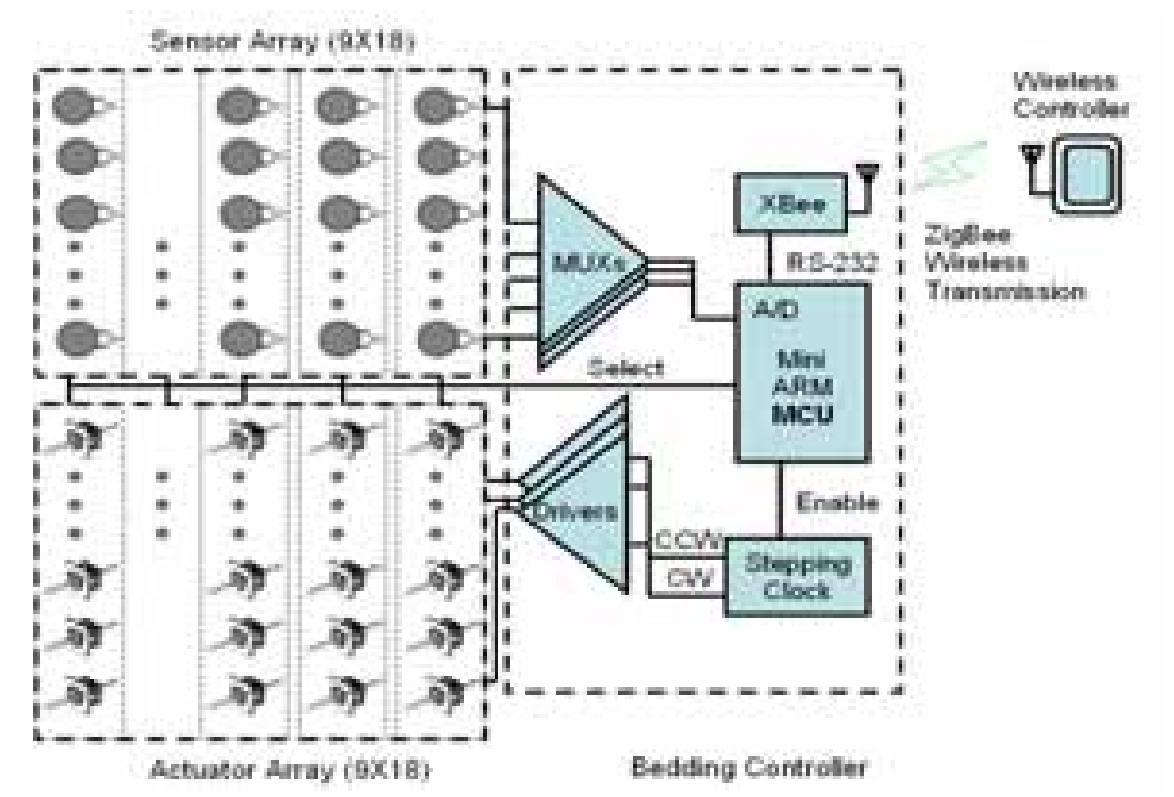

Figure 1. Proposed sensor-based mattress architecture.

## III. Key Components and Designed Modules

Some key components and designed modules are given in this section.

### A. Flexible pressure sensors

The pressure sensor is an ultra-thin flexible force sensing sensor Uneo GS25-100N developed by UCCTW Co., suitable for a wide range of human-machine interface measurement from a gentle finger touch to full-body weighing scales. The sensor technology employs the latest advances in piezo-resistive polymer composite processing and printing-based micromachining technology to enable simple and high-quality linear output in the form of variable conductance (inverse of resistance) that is proportional to the input force from 0 to 100N [5]. We used an operational amplifier in an inverting configuration to measure a voltage output that varies linearly with respect to force input, as shown in Fig. 2, where Rs is an equivalent resistance of pressure sensor.

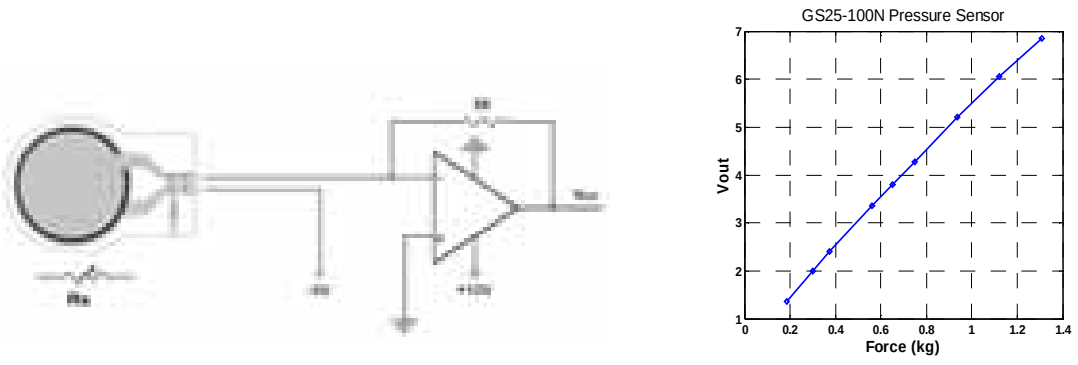

Figure 2. The circuit to measure the GS25-100N sensor and a voltage output with respect to force input.

### B. Micro-type Actuator

We choose in this research the non-captive linear actuator, A43F4AB-24-001, developed also by HaydonKerk incorporation, as the mattress support elements. This stepping motor is designed by replacing rotary motors which generate linear movement and can be run on the stepping clock driven as shown in Fig. 3(a). The movement length is controlled by Enable signal duration. However, one phase must be energized through a properly selected capacitor, e.g., 1 μF in this research, as shown in Fig. 3(b). The linear speed is setting to 10 mm per second and only one synchronous speed. Virtually instantaneous starting and stopping characteristics are one of the advantages of the synchronous stepping motors [6].

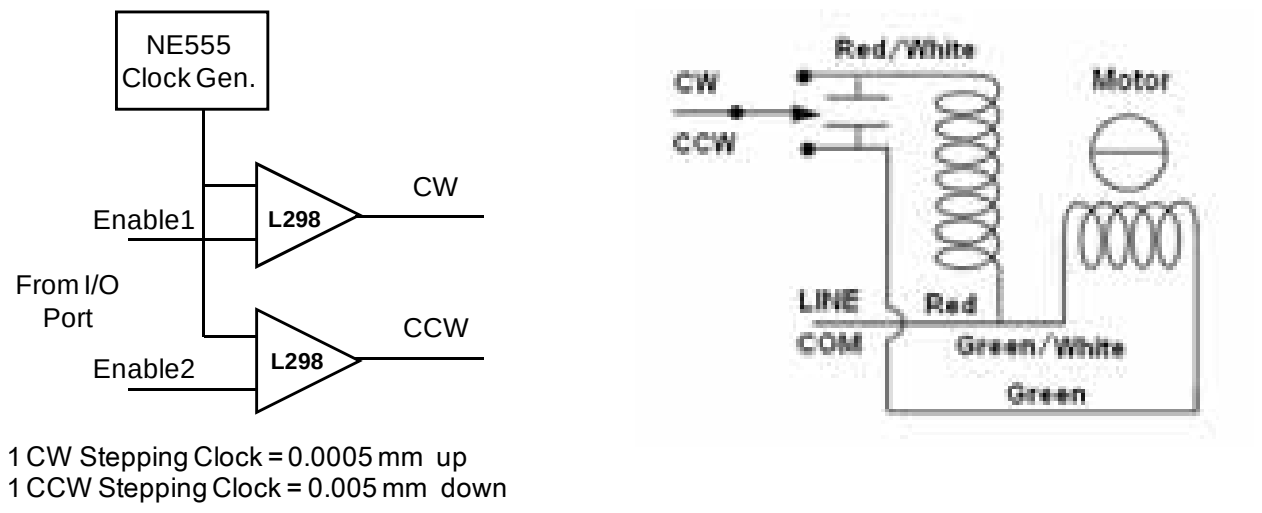

Figure 3. Circuits of stepping clock and motor driving.

### C. The Pressure Sensor and Actuator (PSA) Module

To make the pressure sensor and the actuator work compact, we combine an actuator, a pressure sensor, and a measuring and driving circuit board to form a pressure, sensor and actuator (PSA) module shown as Fig. 4. In Fig. 4, a pressure sensor is attached to the cap atop one end of the non-captive shaft of actuator as well as a circuit board is protected by a can with a diameter 82 mm. Sensed signal lines pass through guided track to control board. On this board, we have pressure signal amplifier, actuator driving circuit, top and bottom limiters, and some interface logical circuits. All signals transfer to or receive from a bedding controller via a 12-pin connector.

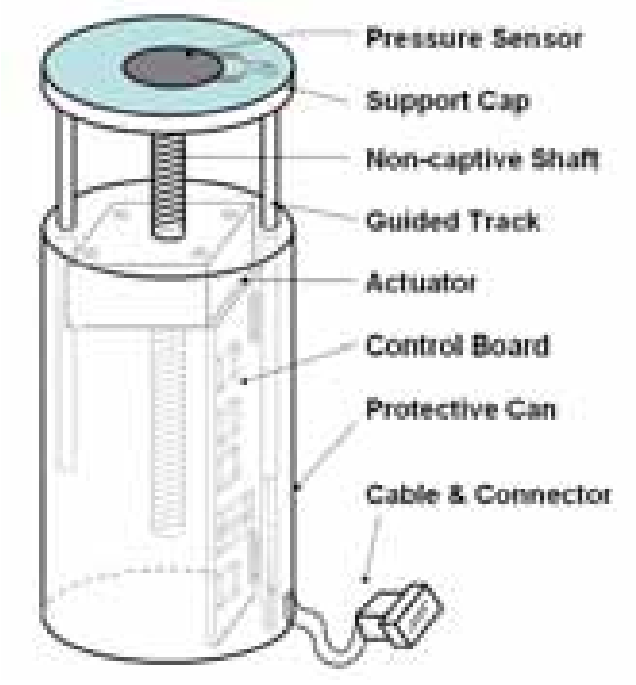

Figure 4. The pressure sensor and actuator (PSA) Module

### D. Bedding Controller

The bedding controller consists of MCU core and electronics. The electronics connects to PSA module, gather the data from the sensors and drive actuators to support the body respectively. The MCU core is a Friendly ARM-based Mini 2440 single-board computer (SBC) with plenty of I/O. Bedding controllers are designed to process pressure readings, generate pressure distribution mapping, calculate uniform support from the mapping, and control actuators in support layer of the mattress. With the provided OS and tools, C++ is used to write application software. After function verified, application software is embedded into this SBC. Then, by the embedded software, the proposed mattress can provide body-contouring uniform support.

### *E. Wireless Controller*

Wireless controller is also a Friendly ARM mini2440, a small single-board computer with a 3.5" touch screen (240x320 pixel), designed for bridging the user and the proposed mattress.

## IV. SUPPORT ALGORITHMS

The support algorithms of the proposed mattress will be addressed in this section. When a human body lies down on a mattress, as shown in Fig. 5(a), the pressure reading at each point on the mattress can be stored in registers of Mini 2440 and a pressure distribution mapping will be generated. Besides, the Mini 2440 controller can also calculate the weight of the body by summing up the acquired pressure value. If the calculated body weight is in the range from 20 to 180 Kg, the operation of uniform support function of the proposed mattress is activated. The average weight can be calculated from the total weight divided by the total number of points these have been pressured. Then according to the different of average weight and pressure of each point, the relative actuator is driven up or down to increase or decrease pressure till all sensed voltage signals of the pressed sensors are equal to uniform support value.

As shown in Fig. 5(b), there are some part of mapping being discontinued, such as neck, low back, and knee, where the body on the mattress with less pressure. In other words, mattress support zero there. By the aid of the mathematical morphology algorithm, we can bridge this discontinuity and define three modes of support firmness such as standard, soft, and medium firmness, as described follows.

### *A. Standard Firmness*

When a human body lied down on a mattress shown in Fig. 5(a), the pressure distribution mapping sensed from pressure sensors array after given a threshold value, for example, 0.05 Kg, can be translated as 9 × 18 pixels binary mapping as shown in Fig. 5(b). Assume i-th row and j-th column pressure sensor reading is $W_{ij}$, then the weight of full body is determined by

$$W = \sum_{i=0}^{M-1}\sum_{j=0}^{N-1} W_{ij} \quad , \; M=18 \;, \; N=9 \qquad (1)$$

Fig. 5(b) can be translated to binary table as shown in Fig 5(c) by the following equation.

$$n_{ij} = \begin{cases} 1 & \text{i f } W_{ij} \geqq 0.05\text{Kg, otherwise} \\ 0 & \end{cases} \qquad (2)$$

where $n_{ij}$ is a scaling value between 0 and 1 of i-th row and j-th column of binary table.

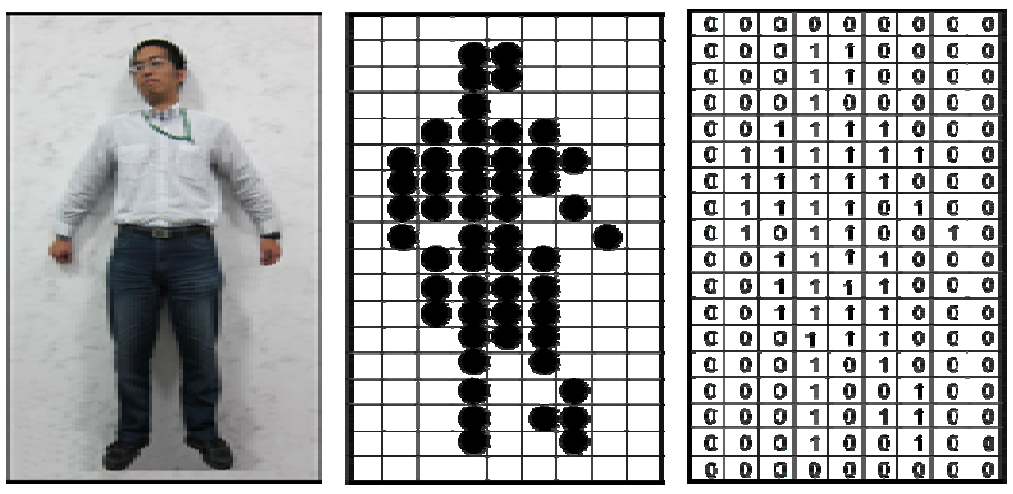

Figure 5. (a) A human body on mattress, (b) Binary mapping of pressure distribution, (c) Binary table of (b).

Define N as the total number of pressured sensors. Then, N is determined by

$$N = \sum_{i=0}^{M-1}\sum_{j=0}^{N-1} n_{ij} \quad , \; M=18 \;, \; N=9 \qquad (3)$$

Define $W_{av}$ is the uniform support for the body. Then, $W_{av}$ is determined by

$$W_{av} = \frac{W}{N} \qquad (4)$$

For example, assuming the weight of the body in Fig. 5(a) is 80 Kg, then $W_{av}$ = 80Kg / 53 = 1.51Kg, i.e. if pressure is less than support force actuator shaft is driven up (CW) to increase pressure sensor reading. Otherwise, the actuator shaft is driven down (CCW) to decrease pressure sensor reading. When pressure reading and uniform support value are balance, the operation is stop.

Let $D_{ij}$ be the different of pressure reading and uniform support value, then the actuator is driven by following equation.

$$D_{ij} = W_{ij} - W_{av} = \begin{cases} \text{CCW} & \text{if } D_{ij} \geqq 0.05\text{Kg}, \\ \text{STOP} & \text{if } 0.05\text{Kg} \geqq D_{ij} \geqq -0.05\text{Kg}, \\ \text{CW} & \text{if } D_{ij} \leqq -0.05\text{Kg} \end{cases} \qquad (5)$$

Finally, each point of the human body is uniformly supported.

### *B. Soft firmness*

For getting more comfortable sleep with full body uniform support, there are two operators, dilation and closing, in the area of mathematic morphology can enlarge the boundary of the human body mapping. In general, dilation can expand the boundaries of image and tends to smooth concavities (joining disparate elements in an image).

This enlarged body boundaries and increase support nodes, and reduce support value, as shown in Fig. 6(b).

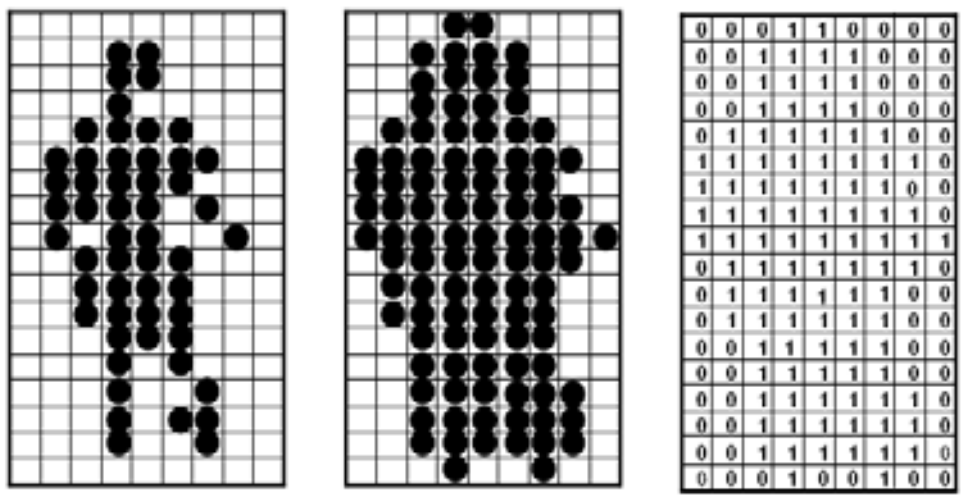

Figure 6. (a) Binary mapping of pressure distribution,(b) Binary mapping after dilation operation,(c) Binary table of (b).

Then, $N = \Sigma\Sigma\, n_{ij} = 101$. The uniform support value of each point is determined by $W_{av} = 80Kg / 101 = 0.79Kg$ 。

### C. *Medium firmness*

Closing is another morphological operation that derived from the basic operators of dilation and erosion. In the case of closing, we dilate first and then erode. Closing is similar in some ways to dilation in that it tends to enlarge the boundaries of image. Fig. 7(a) shows the binary mapping of pressure distribution described as above, Fig. 7(b) shows the binary mapping after closing, and Fig. 7(c) shows the binary table of Fig. 7(b).

Then, $N = \Sigma\Sigma\, n_{ij} = 63$. The uniform support value of the mattress is determined by $W_{av} = 80Kg / 63 = 1.27Kg$ .

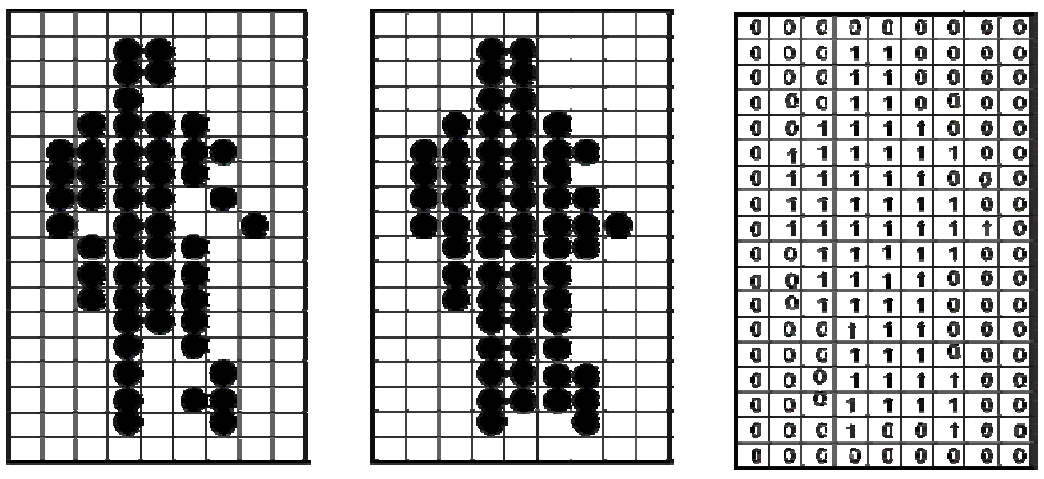

Figure 7. (a) Binary mapping of pressure distribution, (b) Binary mapping after closing operation, (c) Binary table of (b).

## V. System Integration and Testing

After all test processes on the small bed are verification complete, the whole system is realized on an enlarged single bed, as shown in Fig. 8. The size of enlarged single bed is 108 × 188 cm and we use 9 × 18 PSA modules to complete this mattress active support. We also use a personal computer (PC) as a development platform then embedded into a FriendlyARM of bedding controller buried in foam layer. All supporting operations can be down within 20 seconds after the start icon is touched.

## VI. Conclusion

Aimed to the problems of innerspring mattresses that cannot allow body-contouring uniform support and foam mattresses that cannot provide everybody equal comfort from the same mattress, we proposed in this paper a method of pressure sensor-based mattress development for body-contouring uniform support. In addition, we also implement the architecture with an enlarged single mattress. Under the proposed architecture, the new mattress performs a significantly improve for high quality and ultimate conformity sleep environment.

## Acknowledgment

The authors would like to thank the MOEA of Taiwan for financial support of the research (under the SBIR program of ODERBRU Co., Ltd., Project No. 1P1001049).

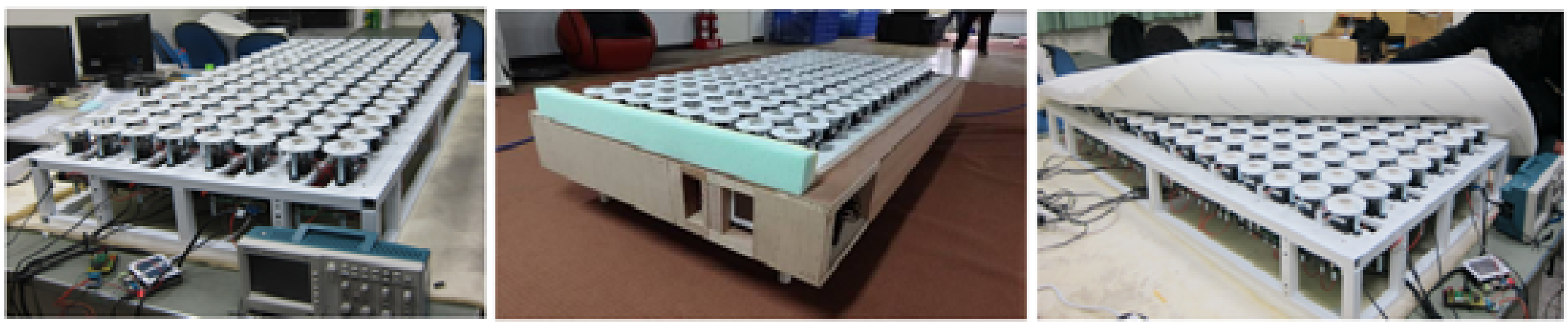

Figure 8. Photos of realized system on an enlarged single bed.